\documentclass{ws-procs975x65}

\begin{document}

\title{GENERALIZED ENTANGLEMENT IN STATIC AND
\break DYNAMIC QUANTUM PHASE TRANSITIONS }

\author{Shusa Deng and Lorenza Viola$^*$}

\address{Department of Physics and Astronomy, Dartmouth College,\\
Hanover, NH 03755, USA\\
%E-mail: Shusa.Deng@Dartmouth.edu \\
$^*$E-mail: Lorenza.Viola@Dartmouth.edu\\
%www.university\_name.edu
}

\author{Gerardo Ortiz}

\address{Department of Physics, Indiana University,\\
Bloomington, IN 47405, USA\\
E-mail: ortizg@indiana.edu}

\begin{abstract}
We investigate a class of one-dimensional, exactly solvable
anisotropic XY spin-$1/2$ models in an alternating transverse magnetic
field from an entanglement perspective.  We find that a physically
motivated Lie-algebraic {\em generalized entanglement} measure
faithfully portraits the static phase diagram -- including second- and
fourth-order quantum phase transitions belonging to distinct
universality classes.  In the simplest time-dependent scenario of a
slow quench across a quantum critical point, we identify parameter
regimes where entanglement exhibits {\em universal dynamical scaling}
relative to the static limit.
\end{abstract}


\bodymatter

\section{Introduction}\label{Intro}

Developing methodologies for probing, understanding, and controlling
quantum phases of matter under a broad range of equilibrium and
non-equilibrium conditions is a central goal of condensed-matter
physics and quantum statistical mechanics.  Since novel forms of
matter tend to emerge in the deep quantum regime where thermal effects
are frozen out, a key prerequisite is to obtain an accurate
theoretical understanding of zero-temperature {\em quantum phase
transitions} (QPTs)~\cite{Sachdev}.  Aside from its broad conceptual
significance, such a need is heightened by the growing body of
experimental work which is being performed at the interface between
material science, quantum device technology, and experimental
implementations of quantum information processing (QIP).  Following
the experimental realization of the Bose-Hubbard model in a confined
${}^{87}$Rb Bose-Einstein condensate and the spectacular observation
of the superfluid-to-Mott-insulator QPT~\cite{Greiner}, ultracold
atoms are enabling investigations into strongly interacting many-body
systems with an unprecedented degree of control and flexibility --
culminating in the observation of topological defects in a rapidly
quenched spinor Bose-Einstein condensate~\cite{Dan}.  Remarkably, the
occurrence of a QPT influences physical properties well into the
finite-temperature regime where real-world systems live, as vividly
demonstrated by the measured low-temperature resistivity behavior in
heavy-fermion compounds~\cite{Gegenwart}.

From a theoretical standpoint, achieving as a complete and rigorous
quantum-mechanical formulation as desired is hindered by the
complexity of quantum correlations in many-body states and dynamical
evolutions.  Motivated by the fact that QIP science provides, first
and foremost, an organizing framework for addressing and quantifying
different aspects of ``complexity'' in quantum systems, it is natural
to ask: Can QIP concepts and tools contribute to advance our
understanding of many-body quantum systems?  In recent years, {\em
entanglement theory} has emerged as a powerful bridging testbed for
tackling this broad question from an information-physics
perspective. On one hand, entanglement is intimately tied to the
inherent complexity of QIP, by constituting, in particular, a {\em
necessary} resource for computational speed-up in pure-state quantum
algorithms~\cite{speedup}.  On the other hand, critically reassessing
traditional many-body settings in the light of entanglement theory has
already resulted in a number of conceptual, computational, and
information-theoretic developments.  Notable advances include
efficient representations of quantum states based on so-called {\em
projected entangled pair states} \cite{states}, improved
renormalization-group methods for both static 2D and time-dependent 1D
lattice systems \cite{dmrg}, as well as rigorous results on the
computational complexity of such methods and the solvability
properties of a class of generalized mean-field Hamiltonians
\cite{complex}.

In this work, we focus on the problem of characterizing quantum
critical models from a {\em Generalized Entanglement} (GE) perspective
\cite{ge1,Somma}, by continuing our earlier exploration with a twofold
objective in mind: first, to further test the usefulness of GE-based
criticality indicators in characterizing static quantum phase diagrams
with a higher degree of complexity than considered so far (in
particular, multiple competing phases); second, to start analyzing
time-dependent, non-equilibrium QPTs, for which a number of
outstanding physics questions remain.  In this context, special
emphasis will be devoted to establish the emergence and validity of
{\em universal scaling laws for non-equilibrium observables}.

\section{Generalized Entanglement in a Nutshell}

\subsection{The need for GE}

Because a QPT is driven by a purely quantum change in the many-body
ground-state correlations, the notion of entanglement appears
naturally suited to probe quantum criticality from an
information-theoretic standpoint: What is the structure and role of
entanglement near and across criticality?  Can appropriate
entanglement measures detect and classify quantum critical points
(QCPs) according to their universality properties?  Extensive
investigations have resulted in a number of suggestive results, see
e.g. Ref. \refcite{fazio} for a recent review. In particular, pairwise
entanglement, quantified by so-called {\em concurrence}, has been
found to develop distinctive singular behavior at criticality in the
thermodynamic limit, universal scaling laws being obeyed in both 1D
and 2D systems. Additionally, it has been established that the
crossing of a QCP point is typically signaled by a logarithmic
divergence of the {\em entanglement entropy} of a block of nearby
particles, in agreement with predictions from conformal field
theory. While this growing body of results well illustrates the
usefulness of an entanglement-based view of quantum criticality, a
general theoretical understanding is far from being reached.  With a
few exceptions, the existing entanglement studies have focused on
analyzing how (i) {\em bipartite} quantum correlations (among two
particles or two contiguous blocks) behave near and across a QCP under
the assumption that the underlying microscopic degrees of freedom
correspond to (ii) {\em distinguishable} subsystems (iii) {\em at
equilibrium}.

GE provides an entanglement framework which is uniquely positioned to
overcome the above limitations, while still ensuring consistency with
the standard ``subsystem-based'' entanglement theory in
well-characterized limits \cite{ge1,Somma,ge2}.  Physically, GE rests
on the idea that entanglement is an {\em observer-dependent} concept,
whose properties are determined by the expectations values of a {\em
distinguished subspace of observables} $\Omega$, without reference to
a preferred decomposition of the overall system into subsystems.  The
starting point is to generalize the observation that standard
entangled pure states of a composite quantum system look mixed
relative to an ``observer'' whose knowledge is restricted to {\em
local} expectation values.  Consider, in the simplest case, two
distinguishable spin-$1/2$ subsystems in a singlet (Bell) state,
\begin{equation}
|\text{Bell}\rangle = \frac{|\uparrow\rangle_A  \otimes 
|\downarrow\rangle_B - |\downarrow\rangle_A \otimes |\uparrow\rangle_B
}{\sqrt{2}} \,,
\label{bell}
\end{equation}
\noindent
defined on a tensor-product state space ${\mathcal H}={\mathcal H}_A
\otimes {\mathcal H}_B$. First, the statement that
$|\text{Bell}\rangle$ is entangled -- $|\text{Bell}\rangle$ cannot be
expressed as $|\psi \rangle_A \otimes |\varphi\rangle_B$ for arbitrary
$|\psi\rangle_A \in {\mathcal H}_A$, $|\varphi\rangle_B \in {\mathcal
H}_B$ -- is unambiguously defined only after a preferred
tensor-product decomposition of ${\mathcal H}$ is fixed: Should the
latter change, so would entanglement in general~\cite{ge2}.  Second,
the statement that $|\text{Bell}\rangle$ is entangled is equivalent to
the property that (either) {\em reduced subsystem state} -- as given
by the partial trace operation, $\rho_A
=\text{Tr}_B\{|\text{Bell}\rangle \langle \text{Bell}|\}$ -- is {\em
mixed}, %that is, non-extremal, 
$\text{Tr}\{\rho_A^2\}
=1/2(1+\sum_{\alpha=x,y,z} \langle \sigma_\alpha^A\rangle^2)<1$, in
terms of expectations of the Pauli spin-$1/2$ matrices
$\sigma_\alpha^A$ acting on $A$.

To the purposes of defining GE, the key step is to realize that a
meaningful notion of a {\em reduced state} may be constructed for any
pure state $|\psi\rangle \in {\mathcal H}$ without invoking a partial
trace, by specifying such a reduced ``$\Omega$-state'' as a list of
expectations of operators in the preferred set $\Omega$.  The fact
that the space of all $\Omega$-states is {\em convex} then motivates
the following \cite{ge1}:

\vspace*{1mm}

{\bf Definition (Pure-state GE).} A pure state $|\psi\rangle \in
{\mathcal H}$ is {\em generalized unentangled relative to $\Omega$ if
its reduced $\Omega$-state is pure, generalized entangled otherwise.}
 
\vspace*{1mm}

For applications to quantum many-body theories, two major advantages
emerge with respect to the standard entanglement definition: first, GE
is directly applicable to both distinguishable and indistinguishable
degrees of freedom, allowing to naturally incorporate
quantum-statistical constraints; second, the property of a many-body
state $|\psi\rangle$ to be entangled or not is independent on both the
choice of ``modes'' (e.g. position, momentum, etc) and the operator
language used to describe the system (spins, fermions, bosons, etc) --
depending only on the observables $\Omega$ which play a distinguished
physical and/or operational role.

\subsection{GE by example}

For a large class of physical systems, the set of distinguished
observables $\Omega$ may be identified with a {\em Lie algebra}
consisting of Hermitian operators, $\Omega \simeq \mathfrak{h}$, 
which generate a corresponding distinguished unitary Lie group via
exponentiation, ${\mathfrak h} \mapsto {\mathcal G}=e^{i {\mathfrak
h}}$.  While the assumption of a Lie-algebraic structure is not
necessary for the GE framework to be applicable \cite{ge1,ge2}, it has
the advantage of both suggesting simple GE measures and allowing a
complete characterization of generalized unentangled states. In
particular, a geometric measure of GE is given by the square length
(according to the trace norm) of the projection of
$|\psi\rangle\langle \psi|$ onto ${\mathfrak h}$:

\vspace*{1mm}

{\bf Definition (Relative purity).} Let $\{ O_\ell \}$, $\ell=1,
\ldots, M,$ be a Hermitian, orthogonal basis for ${\mathfrak h}$,
dim$({\mathfrak h})=M$.  The {\em purity of $|\psi\rangle$ relative to
${\mathfrak h}$} is given by
\begin{equation}
P_{\mathfrak h}(|\psi\rangle) = {\sf K} \sum_{\ell=1}^M \langle \psi|
O_\ell |\psi\rangle^2\,,
\label{hpurity}
\end{equation}
where ${\sf K}$ is a global normalization factor chosen so that 
$0\leq P_{\mathfrak h}\leq 1$.

\vspace*{1mm}

Notice that $P_{\mathfrak h}$ is, by construction, invariant under
group transformations, that is, $P_{\mathfrak h}(|\psi\rangle)=
P_{\mathfrak h}(G|\psi\rangle)$, for all $G \in {\mathcal G}$, as
desirable on physical grounds. If, additionally, ${\mathfrak h}$ is a
semi-simple Lie algebra {\em irreducibly} represented on ${\cal H}$,
generalized unentangled states coincide~\cite{ge1} with {\em
generalized coherent states} (GCSs) of ${\cal G}$, that is, they may
be seen as ``generalized displacements'' of an appropriate reference
state, $|{\sf GCS} (\{ \eta_\ell \}) \rangle = \exp(i \sum_\ell
\eta_\ell O_\ell) |{\sf ref}\rangle$.  Physically, GCSs correspond to
unique ground states of Hamiltonians in ${\mathfrak h}$: States of
matter such as BCS superconductors or normal Fermi liquids are
typically described by GCSs.  While we refer the reader to previous
work \cite{ge1,Somma,ge2} for additional background, we illustrate
here the GE notion by example, focusing on two limiting situations of
relevance to the present discussion.

\subsubsection{Example 1: Standard entanglement revisited}

The standard entanglement definition builds on the assumption of {\em
distinguishable} quantum degrees of freedom, the prototypical QIP
setting corresponding to $N$ local parties separated in real space, and
${\mathcal H}={\mathcal H}_1\otimes\ldots\otimes {\mathcal H}_N$.
Available means for manipulating and observing the system are then
naturally restricted to arbitrary local transformations, which
translates into identifying the Lie algebra of arbitrary local
(traceless) observables, ${\mathfrak h}_{loc}=
{\mathfrak{su}}(\text{dim}({\mathcal H}_1))\oplus \ldots \oplus
{\mathfrak {su}}(\text{dim}({\mathcal H}_N))$, as the distinguished
algebra in the GE approach.  If, for example, each of the factors
${\mathcal H}_\ell$ supports a spin-$1/2$, ${\mathfrak
h}_{loc}=\text{span}\{ \sigma_\alpha^\ell\,; \alpha=x,y,z,
\ell=1,\ldots,N\}$, and Eq.~(\ref{hpurity}) yields
\begin{equation}
{P_{\mathfrak h}}_{loc}(|\psi\rangle)=\frac{1}{N}\sum_{\ell, \alpha}
\langle \psi| \sigma_\alpha^\ell |\psi\rangle^2 =\frac{1}{N}\Big( 
\sum_{\ell} \text{Tr}\rho_\ell^2 -\frac{1}{2}\Big)\,,
\label{locpurity}
\end{equation}
which is nothing but the average (normalized) subsystem purity. Thus,
${P_{\mathfrak h}}_{loc}$ quantifies multipartite subsystem
entanglement in terms of the average bipartite entanglement between
each spin and the rest.  Maximum local purity, ${P_{\mathfrak h}}=1$,
is attained if and only if the underlying state is a pure product
state, that is, a GCS of the local unitary group ${\cal G}_{loc} =
SU(2)_1\otimes \ldots \otimes SU(2)_N$.

\subsubsection{Example 2: Fermionic GE}

Consider a system of indistinguishable spinless fermions able to
occupy $N$ modes, which could for instance correspond to distinct
lattice sites or momentum modes, and are described by canonical
fermionic operators $c_j, c_j^\dagger$ on the $2^N$-dimensional Fock
space ${\mathcal H}_{Fock}$.  Although the standard definition of
entanglement can be adapted to the distinguishable-subsystem structure
associated with a given choice of modes (resulting in so-called ``mode
entanglement''), privileging a specific mode description need not be
physically justified, especially in the presence of many-body
interactions~\cite{Kinder}. These difficulties are avoided in the GE
approach by associating ``generalized local'' resources with {\em
number-preserving} bilinear fermionic operators, which identifies the
unitary Lie algebra ${\mathfrak u}(N)=\text{span}\{ c^\dagger_j c_j\,;
1 \leq i,j\leq N \}$ as the distinguished observable algebra for
fermionic GE. Upon re-expressing ${\mathfrak u}(N)$ in terms of an
orthogonal Hermitian basis of generators, Eq.~(\ref{hpurity}) yields
\begin{equation}
P_{\mathfrak{u}(N)} (|\psi\rangle)=
\frac{2}{N}\hspace*{-1mm}\sum_{j<k=1}^N \Big[ \langle c^\dagger_j c_k
+c^\dagger_k c_j\rangle^2 - \langle c^\dagger_j c_k -c^\dagger_k
c_j\rangle^2 \Big] + \frac{4}{N}\sum_{j=1}^N \langle c^\dagger_j c_j
-1/2 \rangle^2 \,.
\label{fermpurity}
\end{equation}
One may show \cite{Somma} that a many-fermion pure state is
generalized unentangled relative to ${\mathfrak u}(N)$
%($P_{\mathfrak{u}(N)}=1$) 
if and only if it is a single Slater determinant (with any number of
fermions), whereas $P_{\mathfrak{u}(N)}<1$ for any state containing
fermionic GE.  Note that a Bell pure state as in Eq. (\ref{bell})
rewrites, via a Jordan-Wigner isomorphic mapping, in the form
\begin{equation}
|\text{Bell}\rangle =  \frac{|\uparrow\rangle_A  \otimes 
|\downarrow\rangle_B - |\downarrow\rangle_A \otimes |\uparrow\rangle_B
}{\sqrt{2}}=\frac{c^\dagger_1 |\text{vac}\rangle - 
c^\dagger_2 |\text{vac}\rangle }{\sqrt{2}} \,,
\label{bell2}
\end{equation}
in terms of the fermionic vacuum $|{\sf vac}\rangle =
|\hspace*{-1mm}\downarrow\rangle_A \otimes |\hspace*{-1mm}
\downarrow\rangle_B \equiv |\hspace*{-1mm}\downarrow,\downarrow\rangle $.
Thus, while $|\text{Bell}\rangle$ is maximally mode-entangled relative
to the local spin algebra ${\mathfrak{su}}(2)\oplus
{\mathfrak{su}}(2)$, it is ${\mathfrak{u}}(N)$-{\em un}entangled --
consistent with the fact that it is a one-particle state.

\section{Generalized Entanglement and Quantum Critical Phenomena}

\subsection{Static QPTs}

Let us focus in what follows on a class of exactly solvable spin-$1/2$
one-dimensional models described by the following Hamiltonian:
\begin{equation}
H= - \sum_{i=1}^N \left[ \frac{(1+\gamma)}{2} \sigma_x^i
\sigma_x^{i+1} + \frac{(1-\gamma)}{2} \sigma_y^i \sigma_y^{i+1}
\right] + \sum_{i=1}^N \Big( h -(-1)^i \delta \Big)\sigma_z^i,
\label{altH1}
\end{equation}
where periodic boundary conditions are assumed, that is,
$\sigma_\alpha^i \equiv \sigma_\alpha^{i+N}$.  Here, $\gamma \in
[0,1]$, $h \in [-\infty, \infty]$, and $\delta \in [-\infty, \infty]$
are the anisotropy in the XY plane, the uniform magnetic field
strength, and the alternating magnetic field strength,
respectively. For $\delta=0$, the above Hamiltonian recovers the
anisotropic XY model in a transverse field studied in
Ref. \refcite{Somma}, whereas $\delta>0$, $\gamma=1$ corresponds to
the Ising model in a alternating transverse field recently analyzed in
Ref. \refcite{Oleg}.

While full detail will be presented elsewhere \cite{next}, an exact
solution for the energy spectrum of the above Hamiltonian may be
obtained by generalizing the basic steps used in the standard Ising
case \cite{Pfeuty}, in order to account for the existence of a two-site
primitive cell introduced by the alternation.  By first separately
applying the Jordan-Wigner mapping to even and odd lattice sites
\cite{Okamoto}, and then using a Fourier transformation to momentum
space, Hamiltonian (\ref{altH1}) may be rewritten as:
$$ 
H=\sum_{k \in K_+} H_k=\sum_{k \in K_+} \hat{A}_k^{\dag} \hat{H}_k
\hat{A}_k\,,\;\; \;\;K_+=\Big \{\frac{\pi}{N},\frac{3\pi}{N},\ldots,
\Big(\frac{\pi}{2}-\frac{\pi}{N}\Big) \Big\}\,,$$
\noindent 
where $\hat{H}_k$ is a four-dimensional Hermitian matrix, and
$\hat{A}_k^\dag=(a_k^\dag, a_{-k}, b_k^\dag, b_{-k})$ is a vector
operator, $a_k^\dag$ ($b_k^\dag$) denoting canonical fermionic
operators that create a spinless fermion with momentum $k$ for even
(odd) sites, respectively.  \iffalse And
\[ \hat{H}_k = \left ( \begin{array}{cccc}
2*(h+\delta) & 0 & J_k & \Gamma_k \\
0 & -2*(h+\delta) & -\Gamma_k & J_k \\
J_k^\ast & -\Gamma_k^\ast & 2*(h-\delta) & 0 \\
\Gamma_k^\ast & J_k^\ast & 0 & -2*(h-\delta) \end{array}
\right).\]
\noindent with $J_k=-2\cos(k)$, $\Gamma_k=-2i\gamma\sin(k)$.  \fi
Thus, the problem reduces to diagonalizing each of matrices
$\hat{H}_k$, for $k \in K_+$. If $\epsilon_{k,1},
\epsilon_{k,2},\epsilon_{k,3},\epsilon_{k,4}$, with $\epsilon_{k,1}
\leq \epsilon_{k,2} \leq 0 \leq \epsilon_{k,3} \leq\epsilon_{k,4}$
denote the energy eigenvalues of $\hat{H}_k$, then
$$ H_k=\sum_{n=1,\ldots, 4} \epsilon_{k,n} \gamma_{k,n}^\dag
\gamma_{k,n}\,, $$
\noindent 
where $\gamma_{k,n}^\dag, \gamma_{k,n}$ are quasi-particle excitation
operators for mode $k$ in the $n$th band. At $T=0$, the
$\epsilon_{k,1}$ and $\epsilon_{k,2}$ bands are occupied, whereas
$\epsilon_{k,3}$ and $\epsilon_{k,4}$ are empty, thus the ground-state
energy $E_{GS}=\sum_{k \in K_+} (\epsilon_{k,1}+\epsilon_{k,2})$, with
$\epsilon_{k,1} <0, \epsilon_{k,2}\leq 0$.

\begin{figure*}[t]
\begin{center}
\includegraphics[width=6.8cm]{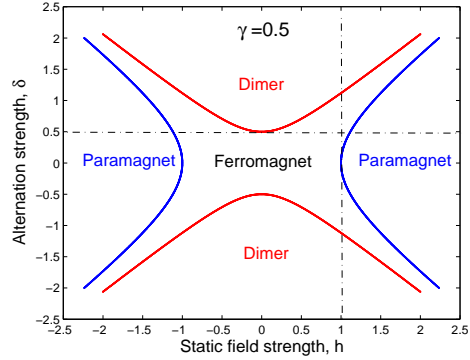}
\caption{Phase diagram of the spin-$1/2$ XY alternating Hamiltonian
given in Eq.~(\ref{altH1}) with $\gamma=0.5$. } \label{fig1}
\end{center}
\end{figure*}

By denoting with $|\text{vac}\rangle$ the fermionic vacuum, and by
exploiting the symmetry properties of the Hamiltonian, the many-body
ground state may be expressed in the form $|\Psi \rangle_{GS}
=\prod_{k\in K^+} |\Psi_k\rangle$, with 
\begin{equation}
|\Psi_k\rangle= \Big( u_k^{(1)} +u_k^{(2)} a_k^\dag
a_{-k}^\dag + u_k^{(3)} b_k^\dag b_{-k}^\dag +
u_k^{(4)} a_k^\dag b_{-k}^\dag +u_k^{(5)} a_{-k}^\dag
b_{k}^\dag  +u_k^{(6)} a_k^\dag a_{-k}^\dag b_k^\dag
b_{-k}^\dag \Big)|\text{vac}\rangle,
\label{GS}
\end{equation}
\noindent 
for complex coefficients determined by diagonalizing $H_k$, with
$\sum_{a=1}^6 |u_k^{(a)}|^2=1$. Since QPTs are caused by
non-analytical behavior of $E_{GS}$, QCPs correspond to zeros
of $\epsilon_{k,2}$. The quantum phase boundaries are determined by 
the following pair of equations: 
$ h^2 = \delta^2+1$; $\delta^2 = h^2+\gamma^2$.  
The resulting {\em anisotropic} ($\gamma >0$) quantum phase diagram is 
showed in Fig.~\ref{fig1} where, without loss of generality, we set 
$\gamma=0.5$. Quantum phases corresponding to disordered 
(paramagnetic, PM) behavior, dimer order (DM), and
ferromagnetic long-range order (FM) emerge as depicted.
In the general case, the boundaries between FM and PM phases, as 
well as between FM and DM phases, are characterized by second-order
broken-symmetry QPTs.  Interestingly, however, $E_{GS}$ develops weak
singularities at
\begin{equation} 
(h_c,\delta_c)  = (0, \delta=\pm\gamma) \,,
\hspace*{1cm} (h_c,\delta_c)  = (\pm 1, \delta=0)\,,
\label{fourth} \end{equation} 
\noindent 
where fourth-order broken-symmetry QPTs occur along the paths
approaching the QCPs (Fig.~\ref{fig1}, dashed-dotted lines). In the
{\em isotropic} limit ($\gamma=0$), an insulator-metal Lifshitz QPT 
occurs from a gapped to a gapless phase, with no broken-symmetry order parameter. For simplicity, we shall primarily focus on {\em gapped} 
quantum phases in what follows, thus $\gamma >0$.
Standard finite-size scaling analysis reveals that new quantum
critical behavior emerges in connection with the alternating
fourth-order QCPs in Eq.~(\ref{fourth})\cite{Oleg}. Thus, in addition
to the usual Ising universality class, characterized by critical
exponents $\nu=1, z=1$, an alternating universality class occurs, with
critical exponents $\nu=2, z=1$.

\begin{figure*}[h]
\centerline{
\includegraphics[width=6.6cm]{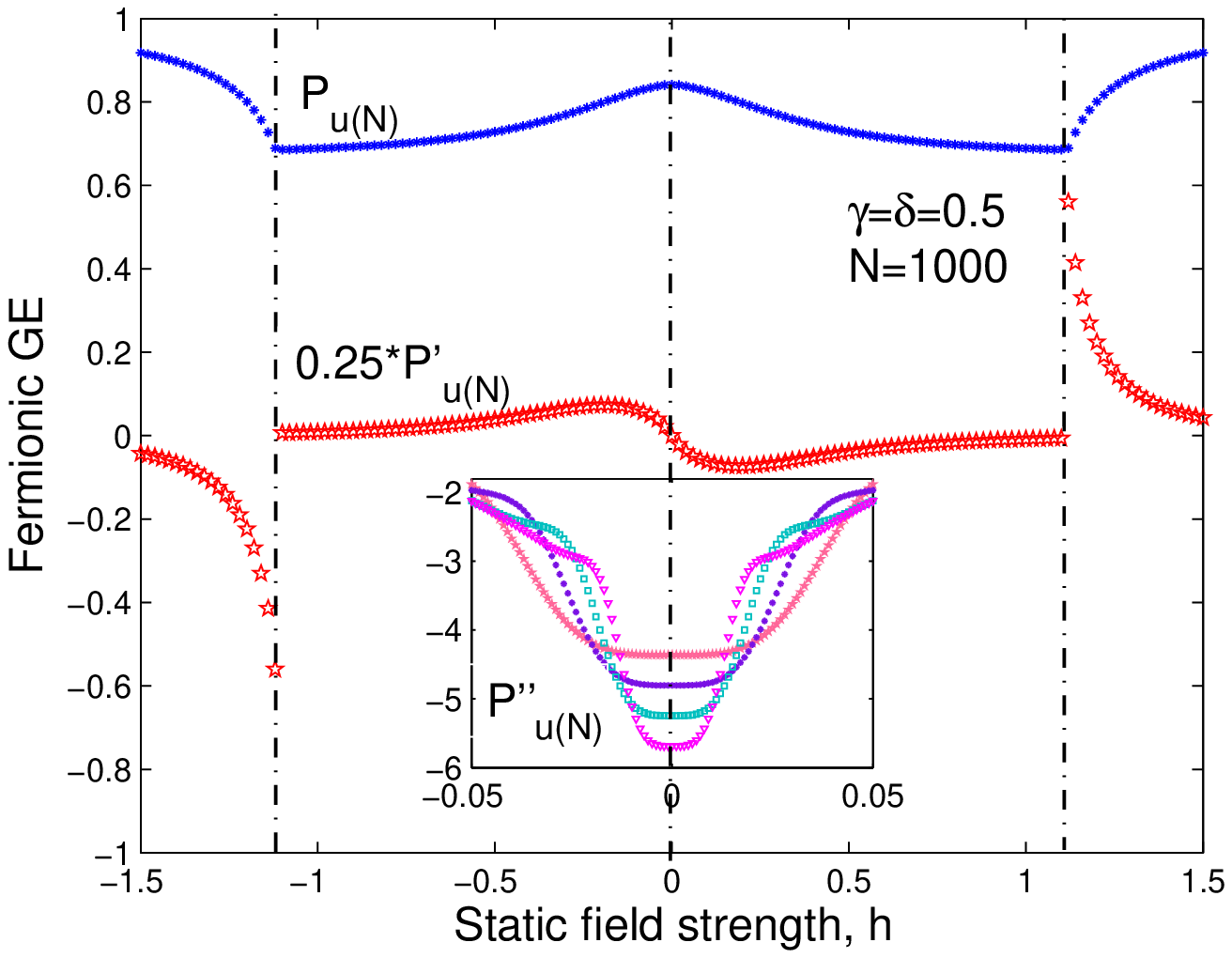}
\includegraphics[width=6.6cm]{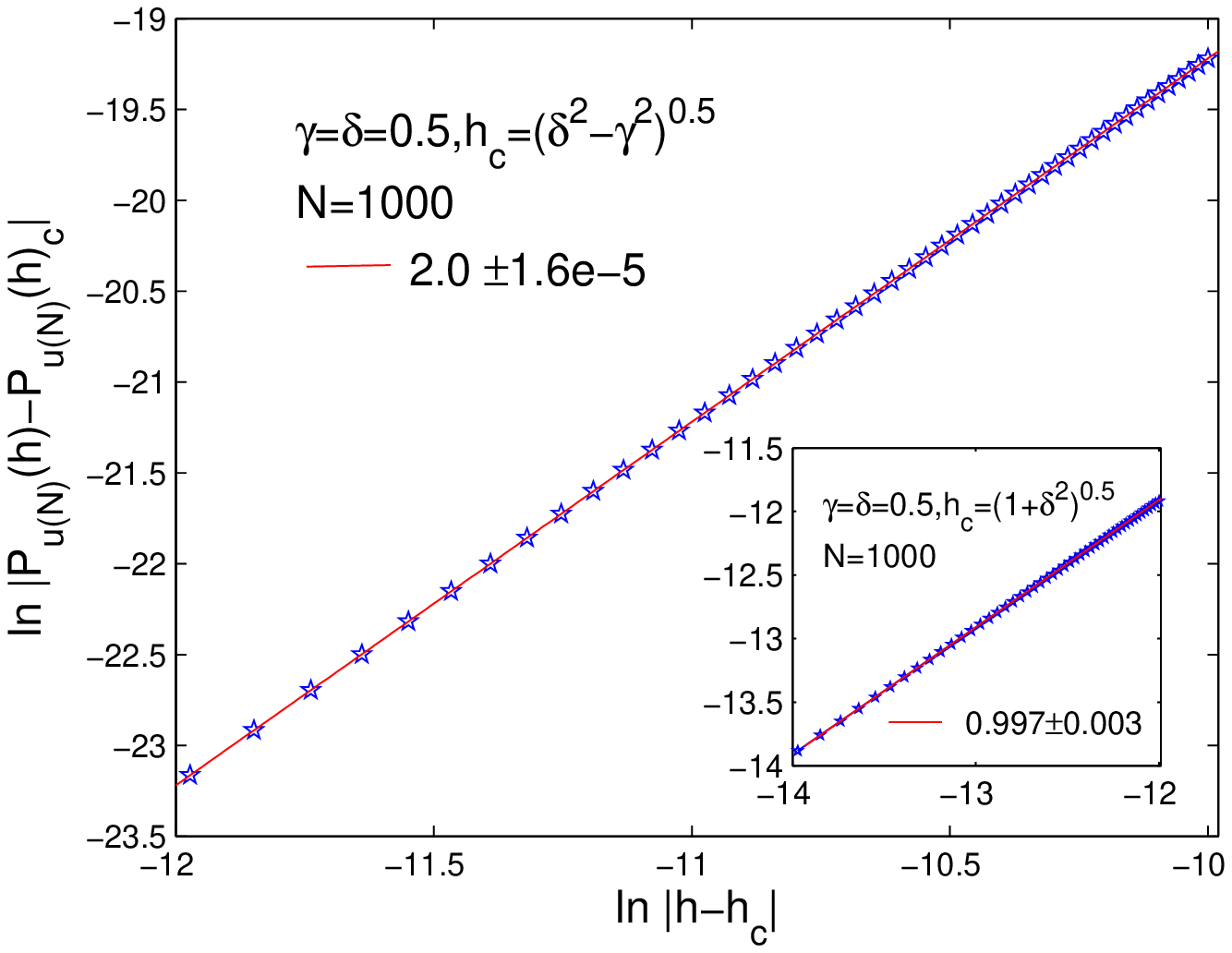}}
\caption{$P_{\mathfrak{u}(N)}$ as a static QPT indicator. Left panel:
Purity and rescaled purity derivative vs magnetic field strength.
Inset: second derivative for $N=1000, 2000, 4000, 8000$ (top to
bottom). Right panel: Determination of $\nu$ for both the alternating
and Ising (inset) universality class.}
\label{fig2}
\end{figure*}

The key step toward applying GE as a QPT indicator is to identify a
(Lie) algebra of observables whose expectations reflect the changes in
the GS as a function of the control parameters.  It is immediate to
realize that Hamiltonian Eq.~(\ref{altH1}), once written in fermionic
language, is an element of the Lie algebra $\mathfrak{so}(2N)$, which
includes {\em arbitrary} bilinear fermionic operators.  As a result,
the GS is always a GCS of $\mathfrak{so}(2N)$, and GE relative to
$\mathfrak{so}(2N)$ carries no information about QCPs. However, the GS
becomes a GCS of the number-conserving sub-algebra $\mathfrak{u}(N)$
in both the fully PM and DM limit.  This motivates the choice of the
fermionic $\mathfrak{u}(N)$-algebra discussed in {\em Example 2} as a
natural candidate for this class of systems.  Taking advantage of the
symmetries of this Hamiltonian, the fermionic purity given in
Eq. (\ref{fermpurity}) becomes:
\begin{eqnarray}
P_{\mathfrak{u}(N)}&=&\frac{8}{N} \sum_{k \in K_+} \bigg\{ \Big[ |\langle
a_k^\dag b_k \rangle| ^2 + |\langle a_{-k}^\dag b_{-k} \rangle|^2 \Big
] \hspace*{4cm}\label{pualter}\\  
 &+&  \frac{4}{N}  \Big[ \langle a_k^\dag a_k
-1/2 \rangle^2 + \langle a_{-k}^\dag a_{-k} -1/2 \rangle^2 + \langle
b_k^\dag b_k -1/2 \rangle^2 + \langle b_{-k}^\dag b_{-k} -1/2
\rangle^2 \Big] \bigg \} \nonumber
\end{eqnarray}
\noindent 
Analytical results for $P_{\mathfrak{u}(N)}$ are only available for
$\delta=0$, where GE sharply detects the PM-FM QPT in the XY model
\cite{Somma}. Remarkably, ground-state fermionic GE still faithfully
portraits the full quantum phase diagram with alternation. First,
derivatives of $P_{\mathfrak{u}(N)}$ develop singular behavior only at
QCPs, see Fig.~\ref{fig2} (left). Furthermore, GE exhibits the correct
scaling properties near QCPs \cite{Somma}.  By taking a Taylor
expansion, $P_{\mathfrak{u}(N)}(h)- P_{\mathfrak{u}(N)}(h_c) \sim
{\xi}^{-1} \sim (h-h_c)^{\nu}$, where $\xi$ is the correlation length,
the static critical exponent $\nu$ may be extracted from a log-log
plot of $P_{\mathfrak{u}(N)}$ for both the Ising and the alternating
universality class, as demonstrated in Fig.~\ref{fig2} (right).

\subsection{Dynamic QPTs}

While the above studies provide a satisfactory understanding of {\em
static} quantum critical properties, {\em dynamical} aspects of
QPTs present a wealth of additional challenges.  To what extent can
non-equilibrium properties be predicted by using equilibrium critical
exponents?  The simplest dynamical scenario one may envision arises
when a single control parameter is slowly changed in time with
constant speed $\tau_q >0$, that is, $ g(t)-g_c={(t-t_c)}/{\tau_q}$,
so that a QCP is crossed at $t=t_c$ ($t_c=0$ without loss of
generality).  The typical time scale characterizing the response of
the system is the {\em relaxation time} $\tau={\hbar}/{\Delta} \sim
{|g(t)-g_c|^{-z\nu}}$, $\Delta$ being the gap between the ground state
and first accessible excited state and $z$ the dynamic critical
exponent \cite{Sachdev}. Since the gap closes at QCPs in the
thermodynamic limit, $\tau$ diverges even for an arbitrarily slow
quench, resulting in a {\em critical slowing-down}. According to the
so-called Kibble-Zurek mechanism (KZM) \cite{Zurek1}, a crossover
between an (approximately) adiabatic regime to an (approximately)
impulse regime occurs at a freeze-out time $-\hat{t}$, whereby the
system's instantaneous relaxation time matches the transition rate,
$$\tau(\hat{t})=|(g(\hat{t})-g_c)/g'(\hat{t})|\,,\;\;\; \;\; \hat{t}
\sim \tau_q^{\nu z/(\nu z +1)}\,,$$ 
\noindent
resulting in a predicted scaling of the final density of excitations
as
\begin{eqnarray}
n(t_F) \sim \tau_q^{-\nu/(\nu z+1)}\,. \label{sKZM}
\end{eqnarray}
\noindent 
While agreement with the above prediction has been verified for
different quantum systems \cite{Jacek}, several key points remain to
be addressed: What are the required physical ingredients for the KZM
to hold? What features of the initial (final) quantum phase are
relevant? How does dynamical scaling reflect into entanglement and 
other observable properties?

\begin{figure*}[h]
\centerline{
\includegraphics[width=6.6cm]{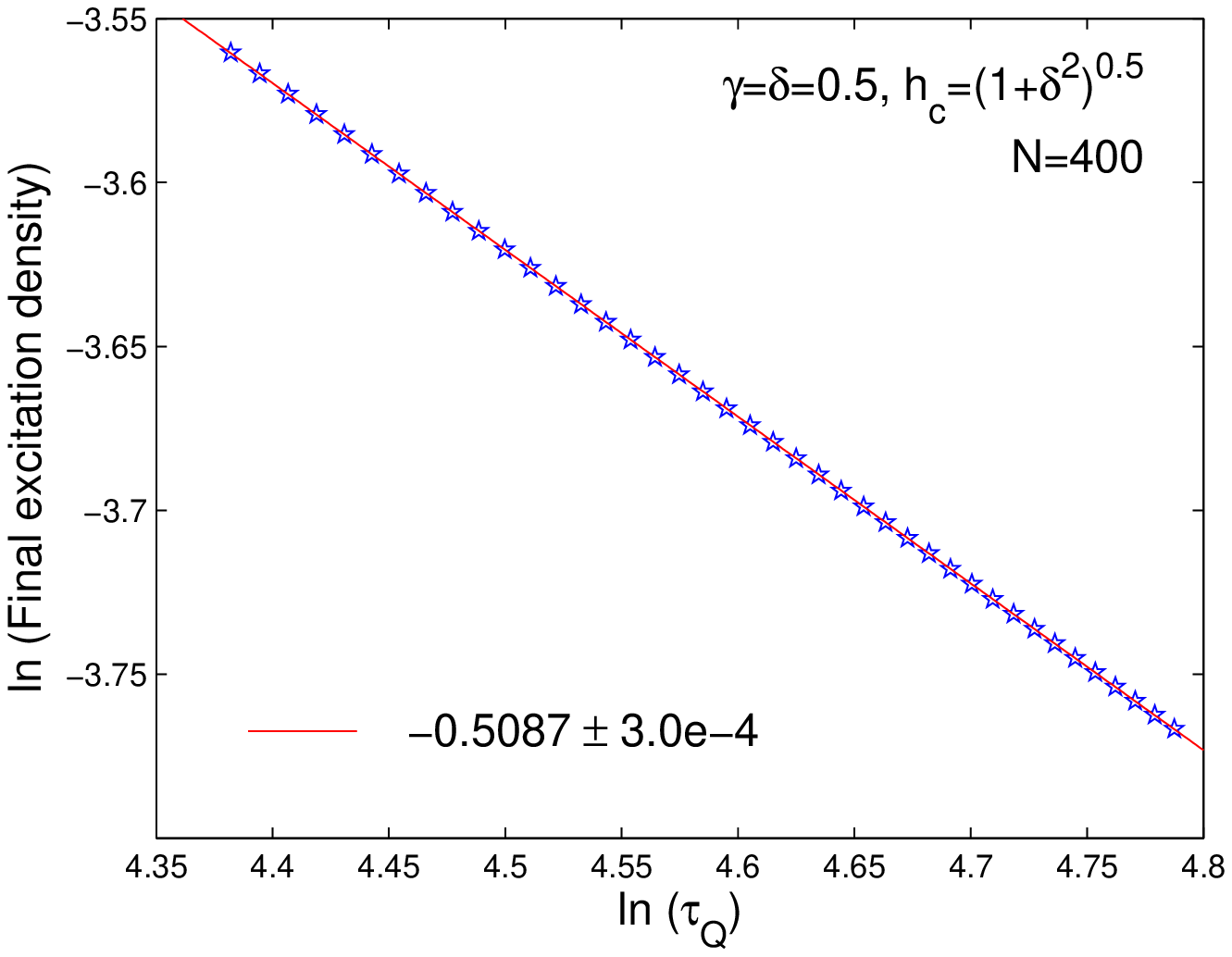}
\includegraphics[width=6.6cm]{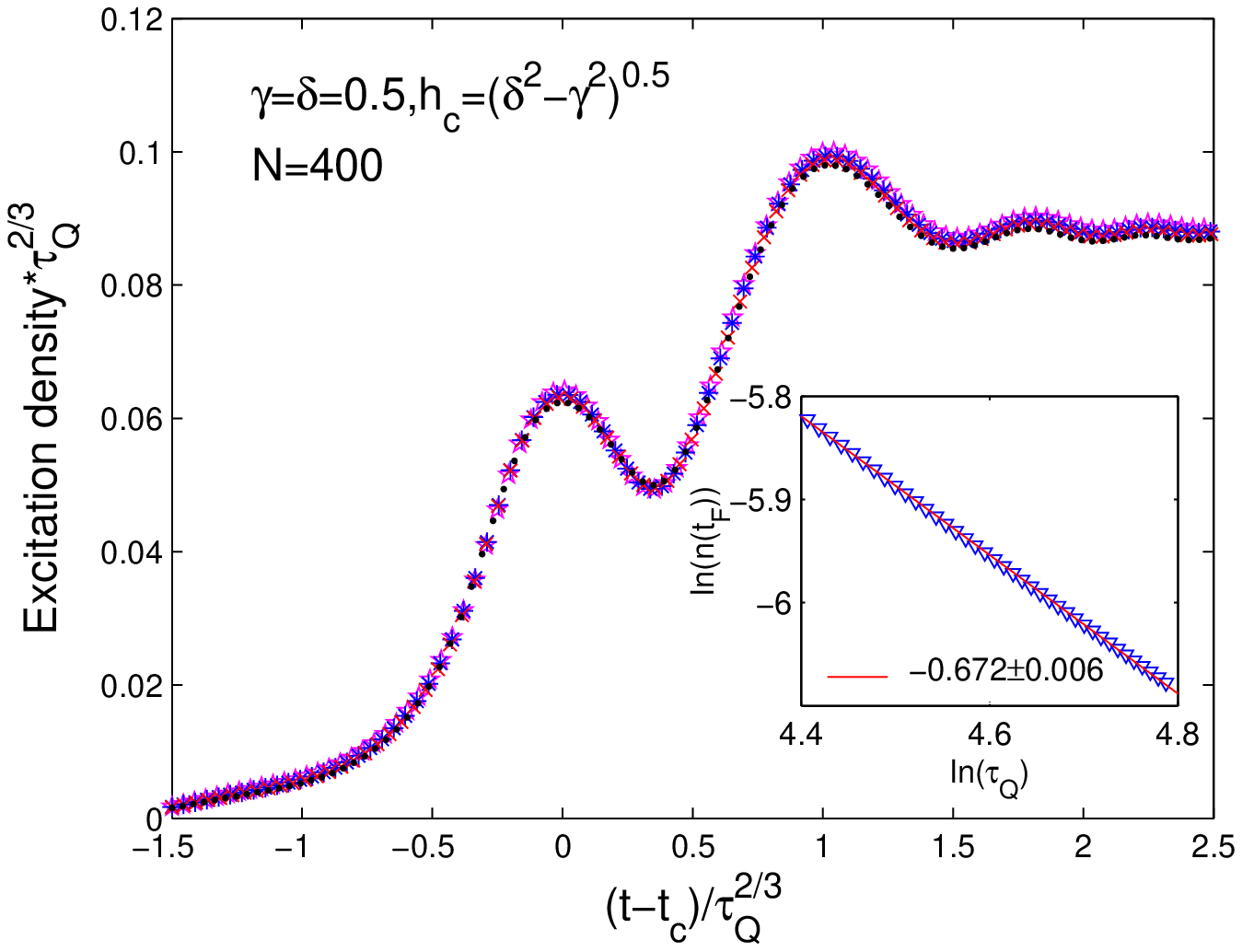}}
\caption{Dynamical scaling of the excitation density. Left panel:
log-log plot for Ising universality class (FM to PM). Right panel: 
alternating universality class (FM to FM), with log-log scaling 
plot in the inset.}
\label{fig3}
\end{figure*}

In our model, the time-evolved many-body state at instant time $t$,
$|\Phi (t) \rangle=\prod_{k \in K^+} |\Phi_k (t) \rangle$, may still
be expressed in the form of Eq. (\ref{GS}) for time-dependent
coefficients $u_k^{(a)} (t)$, $a=1,\ldots, 6$, computed from the
solution of the Schr\"odinger equation, subject to the initial condition 
that $|\Phi(t \rightarrow -\infty)\rangle = |\Psi_{GS}(-\infty)\rangle$.  
The final excitation density is then obtained from the expectation 
value of the appropriate quasi-particle number operator over the 
final state,
$$n(t_F)=\frac{1}{N}\langle \Phi (t_F)| \sum_{k \in K_+} (
\gamma_{k,3}^\dagger \gamma_{k,3} +\gamma_{k,4}^\dagger
\gamma_{k,4})\, | \Phi(t_F)\rangle \,.$$
\noindent
As shown in Fig.~\ref{fig3} (left), the resulting value agrees with
Eq.~(\ref{sKZM}) over an appropriate $\tau_q$-range {\em irrespective
of the details of the QCP and the initial (final) quantum phase}:
$$ n(t_F)^{Ising} \sim \tau_q^{-1/2}\,, \;\;\;\;\;
n(t_F)^{Alternating} \sim \tau_q^{-2/3}\,. $$ 

\begin{figure*}[t]
\begin{center}
\includegraphics[width=7cm]{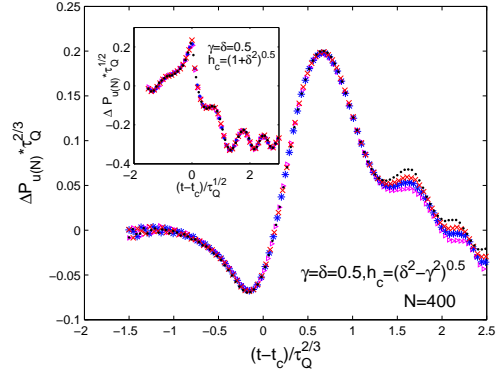}
\caption{Dynamical scaling of $P_{\mathfrak{u}(N)}$ for the
alternating and the Ising (inset) universality class. }
\label{fig4}
\end{center}
\end{figure*}

Remarkably, however, our results indicate that {\em 
scaling behavior holds throughout the entire time evolution} (see
Fig.~\ref{fig3}, right), implying the possibility to express the
time-dependent excitation density as:
\begin{eqnarray*}
n(t)=\tau_q^{-\nu/(\nu z+1)}F\Big(\frac{t-t_c}{\hat{t}}\Big)\,,
\end{eqnarray*}
where $F$ is a universal scaling function. Numerical results support
the conjecture that similar {\em universal dynamical scaling} holds
for arbitrary observables \cite{next}.  In particular, fermionic GE
obeys scaling behavior across the entire dynamics provided that the
amount relative to the instantaneous ground state
$|\Psi(t)\rangle_{GS}$ is considered:
\begin{eqnarray*}
\Delta P_{\mathfrak{u}(N)}(t)\equiv
P_{\mathfrak{u}(N)}(|\Phi(t)\rangle)-P_{\mathfrak{u}(N)}
(|\Psi(t)\rangle_{GS})
=\tau_q^{-\nu/(\nu z+1)}G\Big(\frac{t-t_c}{\hat{t}}\Big)\,,
\end{eqnarray*}
\noindent 
for an appropriate scaling function $G$, see Fig.~\ref{fig4}.

It is important to stress that the above discussion applies to 
control paths which originate and end in {gapped} phases.  
In the isotropic limit $\gamma=0$, we observe no scaling of the 
form Eq.~(\ref{sKZM}) if the system is driven to/from the 
superfluid gapless phase.

\section{Conclusion}

In addition to further demonstrating the usefulness of the GE notion
toward characterizing static quantum critical phenomena, we have
tackled the study of time-dependent QPTs in a simple yet illustrative
scenario. Our analysis points to the emergence of suggestive physical
behavior and a number of questions which deserve to be further
explored.  In particular, while for {\em gapped} systems as considered
here, the origin of the observed universal dynamical scaling is likely
to be rooted in the existence of a well-defined adiabatic (though
non-analytic) limit -- as independently investigated in
Ref. \refcite{Anatoli}, a rigorous understanding remains to be
developed.  We expect that a GE-based perspective will continue to
prove valuable to gain additional insight in quantum-critical physics.

\section*{Acknowledgments}

It is a pleasure to thank Rolando Somma and Anatoli Polkovnikov for
useful discussions and input. Shusa Deng gratefully ackowledges partial
support from Constance and Walter Burke through their {\em Special
Project Fund in Quantum Information Science}.


\begin{thebibliography}{99}

\bibitem{Sachdev} S. Sachdev, {\em Quantum Phase Transitions}
(Cambridge UP, Cambridge, 1999).

\bibitem{Greiner} M. Greiner, O. Mandel, T. Esslinger,
T. W. H\"{a}nsch, and I. Bloch, {\em Nature} {\bf 415}, p. 39 (2002).

\bibitem{Dan} L. E. Sadler, J. M. Higbie, S. R. Leslie,
M. Vengalattore, and D. M. Stamper-Kurn, {\em Nature} {\bf 443},
p. 312 (2006).

\bibitem{Gegenwart} P. Gegenwart {\em et al.}, {\em Phys. Rev. Lett.},
{\bf 89}, p. 056402 (2002).

\bibitem{speedup} R. Jozsa and N. Linded, {\em Proc. Roy. Soc. London
A} {\bf 459}, p. 2001 (2003); G. Vidal, {\em Phys. Rev. Lett.} {\bf
91}, p. 147902 (2003).

\bibitem{states} F. Verstraete and J. I. Cirac, arXiv:
cond-mat/0407066 (2004); {\em Phys. Rev. A} {\bf 70}, p. 060302(R)
(2004).

\bibitem{dmrg} D. Porras, F. Verstraete, and J. I. Cirac, {\em
Phys. Rev. B} {\bf 73}, p. 014410 (2006); G. Vidal, {\em
Phys. Rev. Lett.} {\bf 93}, p. 040502 (2004).

\bibitem{complex} J. Eisert, {\em Phys. Rev. Lett.} {\bf 97},
p. 260501 (2006); R. Somma, H. Barnum, G. Ortiz, and E. Knill, {\em
Phys. Rev. Lett.} {\bf 97}, p. 190501 (2006).

\bibitem{ge1} H. Barnum, E. Knill, G. Ortiz, and L. Viola, {\em
Phys. Rev. A}, {\bf 68}, p. 032308 (2003); H. Barnum, E. Knill,
G. Ortiz, R. Somma, and L. Viola, {\em Phys. Rev. Lett.}, {\bf 92},
p. 107902 (2004).

\bibitem{Somma} R. Somma, G. Ortiz, H. Barnum, E. Knill, L. Viola,
{\em Phys. Rev. A} {\bf 70}, p. 042311 (2004); R. Somma, H. Barnum,
E. Knill, G. Ortiz, and L. Viola, {\em Int. J. Mod. Phys. B} {\bf 20},
2760 (2006).

\bibitem{fazio} L. Amico, R. Fazio, A. Osterloh, and V. Vedral, arXiv:
quant-ph/0703044 (2007).

\bibitem{ge2} L. Viola and H. Barnum, arXiv:quant-ph/0701124 (2007),
and references therein.

\bibitem{Kinder} M. Kindermann, {\em Phys. Rev. Lett.} {\bf 96},
p.  240403 (2006).

\bibitem{Pfeuty} P. Pfeuty, {\em Ann. Phys.} {\bf 57}, p. 79 (1970);
E. Barouch, B. M. McCoy, and M. Dresden, {\em Phys. Rev.  A} {\bf 2},
p. 1075 (1970).

\bibitem{Oleg} O. Derzhko and T. Krokhmalskii, {\em Czech. J. Phys.}
{\bf 55}, p. 605 (2005); O. Derzhko, J. Richter, and T. Krokhmalskii,
{\em Phys.  Rev. E} {\bf 69}, p. 066112 (2004).

\bibitem{Okamoto} K. Okamoto and K. Yasumura, {\em
J. Phys. Soc. Japan} {\bf 59}, p. 993 (1990).

\bibitem{next} S. Deng, G. Ortiz, and L. Viola, in preparation.

\bibitem{Zurek1} W. H. Zurek, U. Dorner, and P. Zoller, {\em
Phys. Rev.  Lett.} {\bf 95}, p. 105701 (2005).

\bibitem{Jacek} J. Dziarmaga, {\em Phys. Rev. Lett.} {\bf 95},
p. 245701 (2005); F. M. Cucchietti, B. Damski, J. Dziarmaga, and
W. H. Zurek, {\em Phys. Rev. A} {\bf 75}, p. 023603 (2007).

\bibitem{Anatoli} A. Polkovnikov, {\em Phys. Rev. B}, {\bf 72},
p. 161201 (2005); A. Polkovnikov and V. Gritsev, cond-mat/0706.0212
(2007).

\end{thebibliography}
\end{document}